\documentclass[aps,prb,showkeys,showpacs]{revtex4}

\usepackage{amsmath}
\usepackage{amsfonts}
\usepackage{amssymb}

\usepackage{graphicx}
\usepackage{dcolumn}
\usepackage{bm}

\begin{document}


\title{Time reparametrization symmetry in spin glass models}

\author{Horacio E. Castillo}

\email[]{castillh@ohio.edu}

\affiliation{Department of Physics and Astronomy, Ohio University,
  Athens, OH, 45701, USA}

\date{\today}

\begin{abstract}
We study the long-time aging dynamics of spin-glass models with
two-spin interactions by performing a Renormalization Group
transformation on the time variable in the non-equilibrium dynamical
generating functional. We obtain the RG equations and find that the
flow converges to an exact fixed point. We show that this fixed point
is invariant under reparametrizations of the time variable. This
continuous symmetry is broken, as evidenced by the fact that the
observed correlations and responses are not invariant under it. We
argue that this gives rise to the presence of Goldstone modes, and
that those Goldstone modes shape the behavior of fluctuations in the
nonequilibrium dynamics.
\end{abstract}

%
%
%
%


\maketitle

\section{Introduction}
\label{sec:intro}

Glassy materials are characterized by very slow dynamics, associated
with a dramatic slowdown of molecular relaxation in structural
glasses, and with a dramatic slowdown of spin relaxation in spin
glasses. This slowdown of the dynamics has been captured in great part
by the results obtained by dynamical mean field theories. In the case
of supercooled liquids, the mean-field mode-coupling
approach~\cite{goetze-MC} has been successful in describing some of
the features of the relaxation. In the case of spin glasses, a
dynamical mean field theory based on a Langevin dynamics for the
spins, examined within a functional integral formulation of the
Martin-Siggia-Rose
approach~\cite{Sompolinsky-Zippelius_relax-dyn-EA-mft_prb-25-6860-1982,
  cugliandolo-kurchan_dynamics-p-spin_prl-173-71-1993,
  cugliandolo-kurchan_out-of-equilibrium-SK_jphysa-27-5749-1994,
  bckm_review97}, has been used to study the long-time relaxation. The
dynamical mean field theory of spin glasses has successfully
captured~\cite{cugliandolo-kurchan_dynamics-p-spin_prl-173-71-1993,
  cugliandolo-kurchan_out-of-equilibrium-SK_jphysa-27-5749-1994,
  bckm_review97} some unusual properties of the spin dynamics,
associated with the lack of equilibration, including the presence of
physical aging and the breakdown of the equilibrium
fluctuation-dissipation relations.

However, mean field theories do not allow direct access to a
description of the fluctuations in the dynamics. It turns out that
fluctuations in the dynamics of glassy systems can in fact be rather
strong, as it has been underlined by the discovery of {\em dynamical
heterogeneities}~\cite{Ediger_review00, Sillescu_review99}. Dynamical
heterogeneities are nanometer-scale regions of molecules rearranging
cooperatively at very different rates compared to the bulk. Recent
studies of material systems near their glass transitions have
uncovered substantial experimental~\cite{Kegel-Blaaderen-science00,
Weeks-Weitz, Weeks-Weitz_prl02, Courtland-Weeks-jphysc03,
Israeloff-AFM-physrev, Israeloff-AFM-nature,
sinnathamby-oukris-israeloff_prl05, wang-song-makse_condmat-0611033,
dauchot-marty-biroli_prl-95-265701, abate-durian_arxiv-07074178} and
simulational~\cite{parisi_jpcb99, Glotzer_simulations, Glotzer-Kob,
Glotzer-Lacevic, parsaeian-castillo_aging-hetdyn} evidence for their
presence.  Various attempts at 
theoretically addressing these strong fluctuations have been made,
involving, among others, the ideas of dynamic
facilitation~\cite{Ritort-Sollich_review-kinetically-constrained_advphys-52-219-2003,
Garrahan-Chandler_geometric-scaling-dynhet_prl-89-035704-2002,
Garrahan-Chandler_kinetically-constrained_pnas-100-9710-2003,
berthier-garrahan_jchemphys-119-4367-2003,
berthier-garrahan_crossovers_pre-68-041201-2003}, the presence of a
``random first order phase
transition''~\cite{xia-wolynes_theory-heterogeneity-supercooled_prl-86-5526-2001,
bouchaud-biroli_adam-gibbs-wolynes-scenario_jchemphys-121-7347-2004,
stevenson-schmalian-wolynes_shapes-rearranging-regions_condmat-0507543},
or the use of diagrammatic methods to carefully re-analyze and extend
mode-coupling
theory~\cite{biroli-bouchaud_upper-critical-dim_epl-67-21-2004,
bouchaud-biroli_dynamic-lenght-scale_condmat-0501668}. However, a
detailed theory that explains the dynamical heterogeneities remains
elusive~\cite{Toninelli-Wyart-Berthier-Biroli-Bouchaud_pre-71-041505-2005}.

Recently, a theoretical framework for the study of fluctuations in the
non-equilibrium dynamics of glassy systems has been
proposed~\cite{ckcc_short-rpg_prl-2002, ccck-prl-numeric, cccki-prb,
chamon-charb-cug-reich-sellito_condmat04}, which is based on the
presence of a Goldstone mode associated with a symmetry under
continuous reparametrizations of the time variable. It was argued
there that the presence of this symmetry could provide an explanation
for many of the dynamical heterogeneity effects observed in various
glassy systems. In Ref.~\cite{ckcc_short-rpg_prl-2002}, a sketch of a
proof for the presence of this symmetry was presented. Earlier work
had uncovered the presence of a restricted version of this symmetry,
for the mean-field dynamical equations of some infinite-range spin
glass models~\cite{
cugliandolo-kurchan_dynamics-p-spin_prl-173-71-1993,
cugliandolo-kurchan_out-of-equilibrium-SK_jphysa-27-5749-1994,
bckm_review97, kennett-chamon-rpg}.

In the present work, we present a detailed proof of the presence of
this symmetry under continuous reparametrizations of the time
variable, for the long time dynamics of a generic spin glass model
with two-spin interactions. The proof is based on using the
Renormalization Group to extract the long time behavior of the
theory. It is somewhat unusual in the sense that we coarse grain {\em
time differences} and not positions. In other words, the degrees of
freedom that are ``integrated over'' are the ones associated with the
``fast'' dynamics, where by ``fast'' we mean fast in time, and not
necessarily in space. 

Although involved in some of its details, our procedure is
conceptually simple. We consider a model for a set of soft spins
on a lattice, which contains only two-spin interactions, with a
zero-mean uncorrelated Gaussian distribution for the spin
couplings. We assume a Langevin-type dynamics for the spins, with a
noise term whose amplitude is controlled by the temperature of the
environment.  We use the functional integral formulation of the
Martin-Siggia-Rose approach to describe the Langevin dynamics. We set
up the calculation by writing the generating functional for the spin
correlations and responses, and find that this generating functional
can be written in terms of a functional integral over an auxiliary
field that depends on two times. We set up the Renormalization Group
procedure by defining a cutoff $\tau_0$ for the time differences. We
increase the cutoff slightly, and integrate over all values of the
auxiliary field that correspond to time differences smaller than this
slightly increased cutoff. This integral is actually a gaussian
integral that can be performed exactly. After integrating over the
``fast'' variables, we rescale all times in such a way that the cutoff
goes back to its original value $\tau_0$. We find that the RG flow
converges to a fixed point, which defines the fixed point generating
functional. Finally, we consider a smooth and monotonously increasing but
otherwise arbitrary reparametrization of the time variable $t \to
s(t)$, which induces a transformation of the sources for the
generating functional. We compute the value of the fixed point generating
functional for those transformed values of the sources, and show that
it is the same as for the original values of the sources. In other
words, the reparametrization of the time variable leaves the fixed
point generating functional invariant.

The rest of the paper is organized as follows:   
in Sec.~\ref{sec:model} we introduce and briefly discuss the spin
model, the Martin-Siggia-Rose formalism for the Langevin spin dynamics
and the assumptions about the nature of the random couplings; 
in Sec.~\ref{sec:Z} we obtain an explicit form for the disorder-averaged 
Martin-Siggia-Rose generating functional, which contains the above
mentioned auxiliary fields that play a central role in the formulation
of the Renormalization Group; 
in Sec.~\ref{sec:RG} we introduce our renormalization group procedure,
associated with coarse graining the {\em time differences\/}, derive
the flow equations for the parameters of the action, and find the
fixed point to which the RG flows; 
in Sec.~\ref{sec:rpg} we derive the central result of this work, i.e.,
we show that the fixed point generating functional is invariant under
reparametrizations of the time variable in the sources; 
and in Sec.~\ref{sec:consequences} we discuss the physical
consequences expected from the presence of this symmetry, which have
already been observed in numerical simulations of spin glasses and
structural glasses, and can also be tested for in confocal microscopy
experiments in colloidal glasses. 
Finally, in Sec.~\ref{sec:summary} we summarize our results.

\section{Model}
\label{sec:model}
We consider a spin-glass hamiltonian containing only two-spin
interactions: 
\begin{equation}
  H_0 = \frac{1}{2} \sum_{r, r'} J_{r r'} \phi_{r} \phi_{r'} 
  + \sum_r  W(\phi_{r}). 
  \label{eq:H-def}
\end{equation}
Here the indexes $r, r'$ label the $N$ possible positions in the
(discrete) lattice, the $\phi_{r}$ are soft spin variables, the $J_{r
r'}$ are the spin coupling constants (satisfying $J_{r r'} = J_{r' r}$
and $J_{r r'} = 0$ for $r = r'$), and the one-spin potential $W(\phi)$
is chosen to control the magnitude of the spin variables.  We assume
that the potential $W(\phi)$ is real, even and analytic at $\phi=0$,
i.e.
\begin{equation}
  W(\phi) = \sum_{p=0}^{\infty} w_p \phi^{2p},  
  \label{eq:W-powers}
\end{equation}  
with $w_p = w^{*}_p \quad \forall p$. For example, for the potential
$W(\phi) = \frac{\lambda}{4} (1 - {\phi}^2)^2$, the coefficients are
$w_0 = \frac{\lambda}{4}$, $w_1 = -\frac{\lambda}{2}$, $w_2 =
\frac{\lambda}{4}$, and $w_p = 0 \quad \forall p>2$.

The Langevin equation for the spin variables for a given realization
$\xi_r(t)$ of the noise reads:
\begin{equation}
  \frac{\partial \phi_r}{\partial t} = 
  - \frac{\partial H}{\partial \phi_r} + \xi_r(t). 
  \label{eq:phi-Langevin}
\end{equation}

We assume, as usual, that the noise is Gaussian distributed and
uncorrelated, with a variance that defines the temperature T of the
heat reservoir:
\begin{equation}
  \langle \xi_{r}(t_1) \xi_{r'}(t_2) \rangle 
  = 2 T \delta_{r, r'} \delta(t_1-t_2), 
  \label{eq:noise-def}
\end{equation}
where the angle brackets $\langle \cdots \rangle$ indicate an average over
the noise distribution. 

We compute the derivatives 
\begin{equation}
  \frac{\partial H_0}{\partial \phi_r} = 
  \sum_{r'} J_{r r'} \phi_{r'} + W'(\phi_r),  
  \label{eq:H-derivative}
\end{equation}
where we have used that for all $r$, $J_{r r} = 0$. 

Then the Martin-Siggia-Rose generating
functional~\cite{MSR-functional-integral_dedominicis-peliti}, averaged
over the realizations of the noise, and incorporating the sources,
reads
\begin{eqnarray}
  \langle Z[\ell,h] \rangle & = &
  \int {\cal D} \phi \, {\cal D} \hat{\phi} \,
       {\cal D} \hat{\varphi} \;
       \exp \left\{ L[\phi,\hat{\phi}] 
       + \sum_r \int_{t_0}^{t_f} \! dt \, \left( 
	 \ell_r(t) \phi_r(t) + i h_r(t) \hat{\phi}_r(t) 
	 \right) \right. 
       \nonumber\\
       && \left. + i \sum_r \hat{\varphi_r} 
       \left\{ \phi_r(t_0) - \varphi_r \right\} 
       \right\}, 
       \label{eq:Z-MSR-general}
\end{eqnarray}
where the notation $\langle (\cdots) \rangle$ indicates the average
over the realizations of the noise. We are considering the time
evolution between times $t_0$ and $t_f$ of the spins $\phi_r(t)$, with
initial conditions given by the $\varphi_r$, i.e. $\forall r :
\phi_r(t_0) = \varphi_r$, and the action is given (in general) by
\begin{eqnarray}
  L[\phi,\hat{\phi}] & = &  
  - i \sum_{r} \int_{t_0}^{t_f} \! dt \,  \hat{\phi}_{r}(t) 
  \left(
    \frac{\partial \phi_r(t)}{\partial t} 
    + \left. \frac{\partial H_0}{\partial \phi_r} \right|_{\{\phi\}}
    - i T \hat{\phi}_r(t) 
    \right).
  \label{eq:MSR-action-general}
\end{eqnarray}
In our case, by using Eq.~(\ref{eq:H-derivative}) we obtain  
\begin{eqnarray}
  L[\phi,\hat{\phi}] & = &  
  -i \sum_{r} \int_{t_0}^{t_f} \! dt \,  \hat{\phi}_{r}(t) \left( 
  \frac{\partial \phi_r(t)}{\partial t} + 
  \sum_{r'} J_{r r'} \phi_{r'} + W'(\phi_r)
  - i T \hat{\phi}_r(t) \right). 
  \nonumber\\
  \label{eq:MSR-action}
\end{eqnarray}

We assume that the disorder is given by an uncorrelated, zero-mean,
gaussian distribution for the couplings, i.e.
\begin{equation}
  {\cal P}\{ J \} = 
  \prod_{r < r'} \left(
    \frac{ \exp\left( {- J^{2}_{r r'}}/{4 \: K_{r r'}} \right) }
	 { ( 4 \pi \: K_{r r'} )^{1/2} } 
	 \right).
  \label{eq:ProbJ-def}
\end{equation}
Here the connectivity matrix $2 K_{r r'} = \overline{J^{2}_{r r'}}$
defines the variances of the random couplings, with the notation
$\overline{( \cdots )}$ denoting an average over the disorder.  The
connectivity matrix $K_{r r'}$ encodes the properties of the
model. For example, in the case of the Edwards-Anderson model, $K_{r
  r'} = K > 0$ for $r,r'$ nearest neighbors and is zero otherwise.

The functional $\overline{\langle Z[\ell,h] \rangle}$ allows the
direct computation of measurable quantities: expectation values,
correlations and responses.  The expectation values and the p-point
correlation functions of the field $\phi_r(t)$ are calculated by
taking derivatives of $\overline{\langle Z[\ell,h] \rangle}$ with
respect to the source $\ell$ coupled to
$\phi$~\cite{MSR-functional-integral_dedominicis-peliti}:
\begin{eqnarray}
\overline{ \langle \phi_{r}(t) \rangle }  & = & \left. \frac{ \delta \overline{ \langle Z[\ell,h] \rangle } }{ \delta
  \ell_{r}(t) } \right\vert_{\ell=0, h=0} \\
C_p(r_1, t_1; r_2, t_2; \cdots; r_p, t_p) & \equiv & \overline{ \langle
\phi_{r_1}(t_1) \phi_{r_2}(t_2) \cdots \phi_{r_p}(t_p) \rangle } 
\nonumber\\ & = & 
\frac{ \delta^{(p)} \overline{ \langle Z[\ell,h] \rangle } }{ \delta
  \ell_{r_1}(t) \ell_{r_2}(t_2) \cdots \ell_{r_p}(t_p) }. 
\label{eq:MSR-phi-correlations}
\end{eqnarray}
Here we have used the fact that the generating functional $\overline{ \langle
Z[\ell,h] \rangle }$ reduces to unity for zero sources, i.e. it
satisfies the condition $\overline{ \langle Z[\ell=0, h=0] \rangle } =1 $.

The effect of (possibly time dependent) external fields: $H = H_0 -
\sum_{r} h_r(t) \phi_r(t)$ can also be probed by computing 
response functions:
\begin{eqnarray}
R(r, t | r', t') & \equiv & 
\frac{ \delta \overline{ \langle \phi_{r}(t) \rangle } }{\delta h_{r'}(t')}
\nonumber\\ & = &
\left. \frac{ \delta \overline{ \langle Z[\ell,h] \rangle } }{\delta \ell_{r}(t) \delta h_{r'}(t')} 
\right\vert_{\ell=0, h=0} \nonumber\\
& = & i \overline{ \langle \phi_{r}(t) \hat{\phi}_{r'}(t') \rangle }, \nonumber\\
\chi(r,t |r',t' ) & \equiv & \int_{t'}^{t}dt'' R(r, t | r', t'').
\label{eq:MSR-responses}
\end{eqnarray}
Here $R(r, t | r', t')$ represents the response to an external field
only present at time $t'$, i.e. a ``delta function in time'', and the
integrated response $\chi((r, t | r', t')$ corresponds to a ``step
field'', i.e. an external field the is ``turned on'' at time $t'$ and
``stays on'' until the time $t$ when the spin is measured. An
important property of response functions is that, by causality, the
response $R(r, t | r', t')$ is zero for $t' > t$. Expectation values,
correlations and responses for {\em one disorder realization\/} can
also be computed by formulas that differ from
Equations~(\ref{eq:MSR-phi-correlations}) and (\ref{eq:MSR-responses})
only in that all disorder averaging is removed.

\section{Disorder averaged generating functional}
\label{sec:Z}
Once the distribution of the
couplings is defined, we can average the disorder-dependent
exponential in the action:
\begin{eqnarray}
  z_0[\hat{\phi},\phi] & \equiv & 
  \overline{\exp\left( -i
    \int_{t_0}^{t_f} dt \sum_{r, r'} J_{r r'} \hat{\phi}_{r}(t)
    \phi_{r'}(t) \right) }
  \nonumber\\
  & = & \prod_{r < r'} \int d J_{r r'} \;
  \frac{ \exp\left( {- J^{2}_{r r'}}/{4 \: K_{r r'}} \right) }
       { ( 4 \pi \: K_{r r'} )^{1/2} } \times \nonumber\\
       & & \qquad 
       \exp\left( J_{r r'} \left\{
	 -i \int_{t_0}^{t_f} \! dt \;
	 \left( \hat{\phi}_{r}(t) \phi_{r'}(t) 
	 + \hat{\phi}_{r'}(t) \phi_{r}(t) \right) 
	 \right\} \right) \nonumber\\
       & = & 
       \exp\left\{ - \sum_{r, r'} \frac{ K_{r r'} }{ 2 } 
       \int_{t_0}^{t_f} dt_1 dt_2 \left( 
	 \hat{\phi}_{r}(t_1) \phi_{r'}(t_1) 
	 \hat{\phi}_{r}(t_2) \phi_{r'}(t_2) 
	 + \hat{\phi}_{r'}(t_1) \phi_{r}(t_1) 
	 \hat{\phi}_{r}(t_2) \phi_{r'}(t_2)
	 \right. \right.
	 \nonumber\\
	 & & \qquad \left. \left. 
	 + \hat{\phi}_{r}(t_1) \phi_{r'}(t_1) 
	 \hat{\phi}_{r'}(t_2) \phi_{r}(t_2)
	 + \hat{\phi}_{r'}(t_1) \phi_{r}(t_1) 
	 \hat{\phi}_{r'}(t_2) \phi_{r}(t_2)
	 \right) \right\}.
       \label{eq:integrate-J}
\end{eqnarray}
We now define the notations
$\phi^{0}_r(t)\equiv \hat{\phi}_r(t)$, $\phi^{1}_r(t)\equiv
\phi_r(t)$, $\bar{0} \equiv 1$, $\bar{1} \equiv 0$, which allow
us to write:
\begin{eqnarray}
  z_0[\hat{\phi},\phi] & = & \exp\left( 
  - \frac{1}{2} \sum_{r, r'} K_{r r'}
  \int_{t_0}^{t_f} dt_1 dt_2 \sum_{a, c = 0}^{1}
  \phi^{a}_{r}(t_1) \phi^{c}_{r}(t_2) 
  \phi^{\bar{a}}_{r'}(t_1) \phi^{\bar{c}}_{r'}(t_2) \right).
  \label{eq:Z0-mu}
\end{eqnarray}
Here we can introduce auxiliary two-time fields $Q^{a c}_{r}(t_1,t_2)$ by
performing a Hubbard-Stratonovich transformation:
\begin{eqnarray}
  z_0[\hat{\phi},\phi] 
  & = & \int {\cal D}Q \; \exp\left(
  - \frac{1}{2} \sum_{r, r'} M_{r r'}
  \int_{t_0}^{t_f} dt_1 dt_2 \sum_{a, c = 0}^{1}
  Q^{a c}_{r}(t_1,t_2) Q^{\bar{a} \bar{c}}_{r'}(t_1,t_2)
  \right. \nonumber \\
  & & \qquad
  \left. 
  + i \sum_{r} 
  \int_{t_0}^{t_f} dt_1 dt_2 \sum_{a, c = 0}^{1}
  Q^{a c}_{r}(t_1,t_2) \phi^{a}_{r}(t_1) \phi^{c}_{r}(t_2) 
  \right),
  \label{eq:Z0-Q}
\end{eqnarray}
where $M_{r r'}$ is the matrix inverse of $K_{r r'}$ and $\int {\cal
D} Q \equiv {\cal N}(M) \int \prod_{r,a,c} \prod_{t_1,t_2}
dQ^{ac}_{r}(t_1,t_2)$. Here ${\cal N}(M) = \left(\det\left((2\pi)^{-1}
M \right) \right)^{1/2} $ is an $M$-dependent normalization factor.  

We're now in a position to write down the disorder-averaged generating
functional for the problem:
\begin{equation}
  {\cal Z}[\ell, h]
  \equiv \overline{\langle Z[\ell, h] \rangle} 
  = \int {\cal D}Q \; 
  \exp\left( - S_K[Q] - S_{nl}[Q,\ell,h] \right), 
  \label{eq:Z-expS-Q}
\end{equation}
where
\begin{eqnarray}
  S_K[Q] & \equiv & 
  \frac{1}{2} \sum_{r, r'}  M_{r r'} \int_{t_0}^{t_f} dt_1 dt_2 
  \sum_{a, c = 0}^{1} Q^{ac}_{r}(t_1,t_2) Q^{\bar{a} \bar{c}}_{r'}(t_1,t_2) 
  \label{eq:S-K-def}
  \\
  S_{nl}[Q,\ell,h] & \equiv & 
  - \ln \int {\cal D}\phi^{0} \, {\cal D}\phi^{1} \, 
  {\cal D}\hat{\varphi} \;
  \exp\left\{ i S_{HS} [Q,\phi^{0},\phi^{1}] + i S_{\mbox{\scriptsize
  spin}}[\phi^{0},\phi^{1}] + i S_{BC}[\phi^{1},\hat{\varphi}] \right.
  \nonumber\\
  & & \left. + \sum_r \int_{t_0}^{t_f} \! dt \, \left(
    \ell_r(t) \phi_r(t) + i h_r(t) \hat{\phi}_r(t) 
    \right) \right\}
  \label{eq:S-nl-def}
  \\
  S_{HS} [Q,\phi^{0},\phi^{1}] & = & 
  \sum_{r} \int_{t_0}^{t_f} dt_1 dt_2 \sum_{a, c = 0}^{1}
  Q^{a c}_{r}(t_1,t_2) \phi^{a}_{r}(t_1) \phi^{c}_{r}(t_2) 
  \label{eq:S-HS-def}
  \\
  S_{\mbox{\scriptsize spin}}[\phi^{0},\phi^{1}] & = & 
  - \sum_{r} \int_{t_0}^{t_f} \! dt \,  \phi^{0}_{r}(t) \left( 
  \frac{\partial \phi^{1}_r(t)}{\partial t} + W'(\phi^{1}_r)
  - i T \phi^{0}_r(t) \right)
  \label{eq:S-spin-def}
  \\
  S_{BC}[\phi^{1},\hat{\varphi}] & = &
  \sum_r \hat{\varphi_r} \left\{ \phi^{1}_r(t_0) - \varphi_r \right\}.
  \label{eqa:boundary_action}
\end{eqnarray}

To simplify the algebra, we take from now on the integration limits as
$t_0 = 0$ and $t_f \to \infty$. By combining Eqs.~(\ref{eq:W-powers})
and (\ref{eq:S-spin-def}), we write $S_{\mbox{\scriptsize spin}}$ in a
way that will allow the RG equations to be put in a simple form:
\begin{eqnarray}
    S_{\mbox{\scriptsize spin}}[\phi^{0},\phi^{1}] & = & 
    - \sum_{r} \int_{0}^{\infty} \! dt \, \left(
    \frac{1}{\Gamma} \phi^{0}_{r}(t) 
    \frac{\partial \phi^{1}_r(t)}{\partial t} 
    + \sum_{a, c = 0}^{1} \gamma^{(2)}_{a c} 
    \phi^{a}_{r}(t) \phi^{c}_{r}(t) 
    + \sum_{p=2}^{\infty} \gamma^{(2p)} \phi^{0}_{r}(t)
    (\phi^{1}_r(t))^{2p-1} 
    \right)
    \nonumber \\
    & & \qquad + \frac{i}{2}
    \int_{0}^{\infty} dt_1 \, dt_2
    \sum_{r, r'} K_{r r'} \sum_{a, c = 0}^{1}
    g^{(4)}(t_1-t_2) \;
    \phi^{a}_{r}(t_1) \phi^{c}_{r}(t_2) 
    \phi^{\bar{a}}_{r'}(t_1) \phi^{\bar{c}}_{r'}(t_2). 
  \label{eqa:RG-general-form}
\end{eqnarray}

\section{Renormalization Group}
\label{sec:RG}
We want to introduce an RG transformation on the {\em time
variables}. Since the construction of the RG transformation is a bit
unusual, we will explain it in detail. We introduce a short-time
cutoff $\tau_0 = 1/\Omega_0$ for the time difference $t_1-t_2$. This
only affects the terms in the action containing an integration over
two time variables, namely $S_K[Q]$, $S_{HS} [Q,\phi^{0},\phi^{1}]$
and $S_{\mbox{\scriptsize spin}}[\phi^{0},\phi^{1}]$. The first two
terms take the following form as a starting point for the RG:
\begin{eqnarray}
  S_K[Q] & = & 
  \frac{1}{2} \sum_{r, r'}  M_{r r'} 
  \int_{ \tau_0 \le |t_1-t_2| \atop 0 \le t_1, t_2 < \infty} 
  dt_1 dt_2 
  \sum_{a, c = 0}^{1} Q^{ac}_{r}(t_1,t_2) Q^{\bar{a} \bar{c}}_{r'}(t_1,t_2), 
  \label{eq:S-K-cutoff}
  \\
  S_{HS} [Q,\phi^{0},\phi^{1}] & = & 
  \sum_{r} 
  \int_{ \tau_0 \le |t_1-t_2| \atop 0 \le t_1, t_2 < \infty} 
  dt_1 dt_2 
  \sum_{a, c = 0}^{1}
  Q^{a c}_{r}(t_1,t_2) \phi^{a}_{r}(t_1) \phi^{c}_{r}(t_2). 
  \label{eq:S-HS-cutoff}
\end{eqnarray}
These terms differ from Eqs.~(\ref{eq:S-K-def}) and
(\ref{eq:S-HS-def}) by the removal of the contributions corresponding
to $|t_1-t_2| < \tau_0$. There are two possible natural assumptions
about how this cutoff is implemented: either we assume (i) that the
contributions for those time pairs is directly removed from $S_K[Q]$
and $S_{HS} [Q,\phi^{0},\phi^{1}]$ without any effects on other terms
in the action, or (ii) that the Hubbard-Stratonovich transformation
performed to obtain Eq.~(\ref{eq:Z0-Q}) is undone for time pairs
$|t_1-t_2| < \tau_0$. These two alternative assumptions lead to
slightly different starting points for the RG flow, but in the end the
flow converges to exactly the same fixed point in both cases. This is
reassuring, in the sense that we expect the properties of the long
time dynamics not to depend on the cutoff procedure. The initial
coefficients for $S_{\mbox{\scriptsize spin}}[\phi^{0},\phi^{1}]$ in
Eq.~(\ref{eqa:RG-general-form}) are given by
\begin{eqnarray}
  \Gamma & = & 1 \nonumber \\
  \gamma^{(2)}_{0 0}& = & -i T \nonumber \\
  \gamma^{(2)}_{0 1}& = & \gamma^{(2)}_{1 0} = 2 \, w_1 \nonumber \\
  \gamma^{(2)}_{1 1}& = & 0 \nonumber \\
  \gamma^{(2p)} & = & 2 \, p \, w_p \quad \forall p \ge 2\\
  g^{(4)}(t_1-t_2) & = & \left\{
  \begin{array}{c@{\quad \mbox{for cutoff procedure} \;}c}
    0 & \mbox{(i)} \\
    {\cal C}_{ |t_1-t_2| < \tau_0 } & \mbox{(ii)}.
  \end{array}
  \right.
  \label{eqa:initial-parameters}
\end{eqnarray}
Here the characteristic function ${\cal C}_{\cal P}$ is defined to be
$1$ if ${\cal P}$ is true and $0$ if ${\cal P}$ is false. 

We now perform an RG transformation on the time variables. We separate
the two-time fields $Q$ into fast modes $Q_{>}$ and slow modes
$Q_{<}$: 
\begin{eqnarray}
Q^{a c}_{>, r}(t_1,t_2) & \equiv & \left\{ 
\begin{array}{c@{\quad \mbox{for} \quad}l}
  Q^{a c}_{r}(t_1,t_2) & \tau_0 \le |t_1-t_2| < b\tau_0 \\
  0 & b\tau_0 \le |t_1-t_2|, 
\end{array}
\right.
\label{eq:Q-fast-def}
\\
Q^{a c}_{<, r}(t_1,t_2) & \equiv & \left\{ 
\begin{array}{c@{\quad \mbox{for} \quad}l}
  0 & \tau_0 \le |t_1-t_2| < b\tau_0 \\
  Q^{a c}_{r}(t_1,t_2) & b\tau_0 \le |t_1-t_2|, 
\end{array}
\right.
\label{eq:Q-slow-def}
\end{eqnarray}
with $b > 1$. 
Clearly, we have
\begin{equation}
  Q^{a c}_{r}(t_1,t_2) = 
  Q^{a c}_{>, r}(t_1,t_2) + Q^{a c}_{<, r}(t_1,t_2),
  \label{eq:Q_sum}\\
\end{equation}
and by inspecting Eq.~(\ref{eq:S-K-cutoff}) we find that 
\begin{equation}
  S_K[Q] = S_K[Q_{>} + Q_{<}] = S_K[Q_{>}] + S_K[Q_{<}]. 
  \label{eq:S_K_sum}
\end{equation}
As our next step, we integrate over the fast variables $Q_{>}$ to obtain 
\begin{eqnarray}
  S_{\Omega/b}[Q_{<}] 
  & = & -\ln \int {\cal D}Q_{>} \;
  \exp\left\{ - S_{\Omega}[Q_{>} + Q_{<}] \right\}
  \nonumber \\
  & = & 
  F_{\Omega} + 
  \frac{1}{2} \sum_{r, r'}  M_{r r'} 
  \int_{ b\tau_0 \le |t_1-t_2| \atop 0 \le t_1, t_2 < \infty} 
  dt_1 dt_2 \sum_{a, c = 0}^{1} 
  Q^{a c}_{<, r}(t_1,t_2) Q^{\bar{a} \bar{c}}_{<, r'}(t_1,t_2) 
  \nonumber \\
  & & 
  - \ln \int {\cal D}\phi^{0} \, {\cal D}\phi^{1} \, 
  {\cal D}\hat{\varphi} \;
  \left\{ \int {\cal D}Q_{>} \exp\left(
  -\frac{1}{2} \sum_{r, r'}  M_{r r'} 
  \int_{ \tau_0 \le |t_1-t_2| < b\tau_0\atop 0 \le t_1, t_2 < \infty} 
  dt_1 dt_2 \sum_{a, c = 0}^{1} 
  Q^{a c}_{>, r}(t_1,t_2) Q^{\bar{a} \bar{c}}_{>, r'}(t_1,t_2)    
  \right. \right. \nonumber \\
    & &  
  \left. \left. + i \sum_{r} 
  \int_{ \tau_0 \le |t_1-t_2| < b \tau_0 \atop 0 \le t_1, t_2 < \infty} 
  dt_1 dt_2
  \sum_{a, c = 0}^{1} Q^{a c}_{> r}(t_1,t_2) 
  \phi^{a}_{r}(t_1) \phi^{c}_{r}(t_2) 
  \right) \right\}
    \nonumber \\
    & & \qquad \times \exp\left(   i \sum_{r} 
  \int_{ b\tau_0 \le |t_1-t_2| \atop 0 \le t_1, t_2 < \infty} 
  dt_1 dt_2
  \sum_{a, c = 0}^{1} Q^{a c}_{<, r}(t_1,t_2) 
  \phi^{a}_{r}(t_1) \phi^{c}_{r}(t_2) 
  + i S_{\mbox{\scriptsize spin}}[\phi^{0},\phi^{1}] 
  + i S_{BC}[\phi^{1},\hat{\varphi}]
  \right)
  \label{eqa:S_integrated-fast}
\end{eqnarray}

The factor $\{ \int {\cal D}Q_{>} \; \exp \{ \cdots \} \}$, which contains
  the integration over the fast modes $Q_{>}$, is actually a gaussian integral,
  which evaluates to
\begin{eqnarray}
  & & \left(\det\left\{(2\pi)^{-1} K\right\}\right)^{{\cal V}(\tau_0,b)/2} 
  \nonumber \\
  & & \times \exp\left( -\frac{1}{2} \sum_{r, r'}  K_{r r'}  
  \int_{ \tau_0 \le |t_1-t_2| < b\tau_0\atop 0 \le t_1, t_2 < \infty} 
  dt_1 dt_2 \sum_{a c = 0}^{1}
    \phi^{a}_{r}(t_1) \phi^{c}_{r}(t_2) 
    \phi^{\bar{a}}_{r'}(t_1) \phi^{\bar{c}}_{r'}(t_2)
    \right),
  \label{eqa:S_gaussian-integral-fast}
\end{eqnarray}
where ${\cal V}(\tau_0,b)$ is proportional to the volume of the
two-dimensional (time) region
where the condition $ \tau_0 \le |t_1-t_2| < b\tau_0 $ holds. In this
expression, the determinant prefactor contributes to the
renormalization of the constant term $F_{\Omega}$, and the argument of
the exponential contributes to the renormalization of the function
$g^{(4)}(t_1-t_2)$.

We now not only perform the rescaling of the fields $Q^{a
  c}_{<,r}(t_1,t_2)$ and the time variable, as it would normally be
  done for an RG procedure, but we also simultaneously rescale the
  $\phi^{a}_{r}(t)$ fields and the sources $\{\ell_r(t),h_r(t)\}$, even
  though those quantities were not subject to the integration of ``fast
  modes'':
\begin{eqnarray}
  Q^{a c}_{<, r}(b t'_1,b t'_2) & = & b^{\lambda^{(2)}_{a c}} \, Q^{' a
  c}_{r}(t'_1, t'_2)
    \label{eqa:rescale_Q} \\
  bt' & = & t 
    \label{eqa:rescale_t} \\ 
  \phi^{a}_{r}(b t') & = & b^{\lambda^{(1)}_{a}} \, \phi^{' a}_{r}(t') 
    \label{eqa:rescale_phi} \\
  \ell_{r}(b t') & = & b^{\lambda_{\ell}} \, \ell'_{r}(t') 
    \label{eqa:rescale_ell} \\
  h_{r}(b t') & = & b^{\lambda_{h}} \, h'_{r}(t').
    \label{eqa:rescale_h} 
\end{eqnarray}

We then get
\begin{eqnarray}
  S'_K[Q'] & = & \left( b^{2 + \lambda^{(2)}_{a c} +
  \lambda^{(2)}_{\bar{a} \bar{c}}} \right)
  \frac{1}{2} \sum_{r, r'}  M_{r r'} 
  \int_{ \tau_0 \le |t'_1-t'_2| \atop 0 \le t'_1, t'_2 < \infty} 
  dt'_1 dt'_2 
  \sum_{a, c = 0}^{1} Q^{'ac}_{r}(t'_1,t'_2) Q^{' \bar{a}
  \bar{c}}_{r'}(t'_1,t'_2)  
  \label{eqa:S'-K} \\
  S'_{HS}[Q',\phi^{'0},\phi^{'1}] & = & \left( b^{2 + \lambda^{(2)}_{a c} +
  \lambda^{(1)}_{a} + \lambda^{(1)}_{c}} \right)
  \sum_{r} 
  \int_{ \tau_0 \le |t'_1-t'_2| \atop 0 \le t'_1, t'_2 < \infty} 
  dt'_1 dt'_2 
  \sum_{a, c = 0}^{1} Q^{'ac}_{r}(t'_1,t'_2) 
  \phi^{'a}_{r}(t'_1) \phi^{'c}_{r}(t'_2). 
  \label{eq:S'-HS}
\end{eqnarray}
 
Since the terms $S_K$ and $S_{HS}$ together represent the 4-spin
interaction that makes the model glassy, we demand that they both
should be marginal under the RG. This leads to the conditions
\begin{eqnarray}
0 & = & 2 + \lambda^{(2)}_{a c} +
  \lambda^{(2)}_{\bar{a} \bar{c}}
  \label{eqa:condition-lambda2} \\
0 & = & 2 + \lambda^{(2)}_{a c} +
  \lambda^{(1)}_{a} + \lambda^{(1)}_{c}.
  \label{eqa:condition-lambda1}
\end{eqnarray}  

The second condition can only be satisfied if $\lambda^{(2)}_{a c}$ is
of the form $\lambda^{(2)}_{a c} = f(a) + f(c)$, where $f(a) \equiv -1
-\lambda^{(1)}_{a}$. Inserting this form into the first condition, it
yields $1 + f(a) + f(\bar{a}) = 0$. At this point we still have
freedom to pick among infinitely many possible solutions to this
equation, each one of them defining a different RG transformation. We
decide to choose the assignment $f(a) \equiv -a$, which leads to 
\begin{eqnarray}
  \lambda^{(2)}_{a c} & \equiv & -a-c 
  \label{eqa:lambda2-result} \\
  \lambda^{(1)}_{a} & \equiv & a-1 = -\bar{a}.
  \label{eqa:lambda1-result}
\end{eqnarray}
The choice of this particular solution is natural if we consider a
Reparametrization Group (RpG)
transformation~\cite{kennett-chamon-rpg}, associated with a
reparametrization $s(t)$ of the time variables,  
\begin{equation}
  \tilde{Q}^{a c}_{r}(t'_1,t'_2) = 
  \left(\frac{\partial s}{\partial t'_1}\right)^{a} 
  \left(\frac{\partial s}{\partial t'_2}\right)^{c} 
  Q^{a c}_{r}(s(t'_1),s(t'_2)), 
  \label{eq:rpg-transf}
\end{equation}
where $a,c \in \{0,1\}$. For the special case of a rescaling of times,
$s(t') = b t'$, Eq.~(\ref{eq:rpg-transf}) reduces to
\begin{equation}
  Q^{a c}_{r}(b t'_1,b t'_2) = b^{-a-c} \, \tilde{Q}^{a
  c}_{r}(t'_1, t'_2),
  \label{eq:rpg-rescaling}
\end{equation}
which is completely analogous to Eq.~(\ref{eqa:rescale_Q}) in the case 
$\lambda^{(2)}_{a c} = -a-c$.  

For the source term, we demand that it should be marginal under the
RG, and obtain the rescaling exponents: 
\begin{eqnarray}
  \lambda_{\ell} & = & -1 -\lambda^{(1)}_{1} = -1,
    \label{eqa:lambda-ell-result} \\
  \lambda_{h} & = & -1 -\lambda^{(1)}_{0} = 0.
    \label{eqa:lambda-h-result} 
\end{eqnarray}
It can be checked that, besides $S_K$, $S_{HS}$ and the source term,
the boundary condition term $S_{BC}$ is also marginal under the RG.

We now consider the effect of the RG transformation on the terms
contained in $S_{\mbox{\scriptsize spin}}[\phi^{0},\phi^{1}]$. The
time derivative term is not affected by the integration over fast
modes, and the rescaling of times and fields introduces the following
rescaling: 
\begin{equation}
  \frac{1}{\Gamma} \rightarrow \frac{1}{\Gamma'} = 
  \frac{b^{1 - \bar{0} - \bar{1}}}{b \Gamma} = \frac{1}{b \Gamma}.
  \label{eq:RG-transf-Gamma}
\end{equation}  
If we now write 
\begin{equation}
  b = e^{\delta l},
  \label{eq:define-scale}
\end{equation}  
we get the RG equation 
\begin{equation}
  \frac{d \Gamma}{d l} = \Gamma.
  \label{eq:RG-eqn-Gamma}
\end{equation}  

Similarly we obtain
\begin{eqnarray}
  \frac{d \gamma^{(2)}_{a c}}{d l} & = & (a+c-1) \; \gamma^{(2)}_{a c},
  \label{eqa:RG-eqn-gamma-2} \\
  \frac{d \gamma^{(2p)}}{d l} & = & 0.
  \label{eqa:RG-eqn-gamma-p} 
\end{eqnarray}

Finally, from Eq.~(\ref{eqa:S_gaussian-integral-fast}), we find that the
 only term to receive a contribution from the integration over the
 fast degrees of freedom is the $g^{(4)}(t_1-t_2)$ term:
\begin{equation}
  g^{(4)}(t_1-t_2) \rightarrow g^{' (4)}(t'_1-t'_2) \; = \;
  b^{2 -\bar{a} -\bar{c} -a -c} 
  \times \left( g^{(4)}(bt'_1-bt'_2) + 
	 {\cal C}_{ \tau_0 \le |bt'_1-bt'_2| < b\tau_0 } \right), 
	 \label{eq:RG-transf-g4}
\end{equation} 
and we observe that for this term the rescaling prefactor evaluates to
unity: $b^{2 -\bar{a} -\bar{c} -a -c} = 1$.

By examining the RG flow of Eqs.~(\ref{eq:RG-eqn-Gamma}),
(\ref{eqa:RG-eqn-gamma-2}), (\ref{eqa:RG-eqn-gamma-p}), and
(\ref{eq:RG-transf-g4}), we find the following fixed point values:
\begin{eqnarray}
  \Gamma^* & = & 0, \infty    \nonumber \\ 
  \gamma^{* (2)}_{00} & = & 0, \infty \nonumber \\ 
  \gamma^{* (2)}_{01} & = & \mbox{any number} \nonumber \\ 
  \gamma^{* (2)}_{10} & = & \mbox{any number} \nonumber \\ 
  \gamma^{* (2)}_{11} & = & 0, \infty \nonumber \\ 
  \gamma^{* (2p)} & = & \mbox{any number} \quad \forall p \ge 2 \nonumber \\ 
  g^{* (4)}(t_1-t_2) & = & {\cal C}_{ |t_1-t_2| < \tau_0 }.
\end{eqnarray}
Since the RG flows of all parameters are uncoupled, the solutions
above can be chosen independently for each parameter. The stability
analysis around the fixed points shows that perturbations of $\Gamma$,
$\gamma^{(2)}_{00}$, and $\gamma^{(2)}_{11}$ are {\em relevant} near
$\Gamma^* = 0$, $\gamma^{* (2)}_{00} = \infty$ and $\gamma^{*
(2)}_{11} = 0$ respectively; and are irrelevant near $\Gamma^* =
\infty$, $\gamma^{* (2)}_{00} = 0$ and $\gamma^{* (2)}_{11} = \infty$
respectively. It also shows that perturbations of $\gamma^{(2)}_{01}$,
$\gamma^{(2)}_{10}$, and $\gamma^{(2p)}$ are marginal around any of
their fixed points. Perturbations of $g^{(4)}(t_1-t_2)$ are always
irrelevant. Therefore, for the set of initial conditions given by
Eq.~(\ref{eqa:initial-parameters}), and for almost any other set of
initial conditions, the RG flows for $\Gamma$, $\gamma^{(2)}_{00}$,
and $g^{(4)}(t_1-t_2)$ converge to their stable fixed points. However,
the parameter $\gamma^{(2)}_{11}$ has a starting value which is
exactly at the unstable fixed point $\gamma^{*(2)}_{11} = 0$, and
stays there through the RG flow. Additionally, the parameters
$\gamma^{(2)}_{01}$, $\gamma^{(2)}_{10}$ and $\gamma^{(2p)}$ do not
flow at all, and stay at their initial values. In summary, the
parameters of $S_{\mbox{\scriptsize spin}}[\phi^{0},\phi^{1}]$ flow to
the fixed point values:
\begin{eqnarray}
  \Gamma^* & = & \infty    \nonumber \\ 
  \gamma^{* (2)}_{00} & = & 0 \nonumber \\ 
  \gamma^{* (2)}_{01} & = & 2 \, w_1 \nonumber \\ 
  \gamma^{* (2)}_{10} & = & 2 \, w_1 \nonumber \\ 
  \gamma^{* (2)}_{11} & = & 0 \nonumber \\ 
  \gamma^{* (2p)} & = & 2 \, p \, w_p \quad \forall p \ge 2 \nonumber \\ 
  g^{* (4)}(t_1-t_2) & = & {\cal C}_{ |t_1-t_2| < \tau_0 }.
\end{eqnarray}
As anticipated above, this result is the same for cutoff procedures
(i) and (ii) (and in fact {\em for any other\/} possible initial value
of $g^{(4)}(t_1-t_2)$). 

The fact that $\Gamma$ flows to infinity
indicates that the derivative term does not appear in the fixed point
action. However, the states of the system at different times are still
coupled by three other terms: $S_K[Q]$, $S_{HS}
[Q,\phi^{0},\phi^{1}]$, and the term proportional to
$g^{(4)}(t_1-t_2)$ in $S_{\mbox{\scriptsize
    spin}}[\phi^{0},\phi^{1}]$. We interpret this to indicate that,
while the time derivative terms may be important for the short time
dynamics, when the short time dynamics is ``integrated over'' and only
the long time dynamics remains, the coupling between different times
is provided only by the terms associated to the spin glass
interactions. This is reminiscent of earlier
mean-field calculations of the aging dynamics of spin glasses, in
which the time derivative terms are negligible at long times, and the
coupling between different times is also provided only by the spin-glass
interaction terms~\cite{
  cugliandolo-kurchan_dynamics-p-spin_prl-173-71-1993,
  cugliandolo-kurchan_out-of-equilibrium-SK_jphysa-27-5749-1994}. In
that context, the time derivative terms break the mean-field version
of time reparametrization invariance, and therefore this invariance is
only valid for very long times, when time derivative terms are
negligible~\cite{ bckm_review97, 
  cugliandolo-kurchan_dynamics-p-spin_prl-173-71-1993,
  cugliandolo-kurchan_out-of-equilibrium-SK_jphysa-27-5749-1994, 
  kennett-chamon-rpg}.

By combining all the results for the RG flow for the various terms in
the action, we find that the action converges to the fixed point:
\begin{eqnarray}
  S_{\mbox{\scriptsize fp}}[Q] & = & F_{\Omega} 
  + \frac{1}{2} \sum_{r, r'}  M_{r r'} 
  \int_{ \tau_0 \le |t_1-t_2| \atop 0 \le t_1, t_2 < \infty} 
  dt_1 dt_2 \sum_{a, c = 0}^{1} 
  Q^{a c}_{r}(t_1,t_2) Q^{\bar{a} \bar{c}}_{r'}(t_1,t_2) 
  \nonumber \\
  & & 
  - \ln \int {\cal D}\phi^{0} \, {\cal D}\phi^{1} \, 
  {\cal D}\hat{\varphi} \;
  \exp\left\{
  i \sum_{r} \int_{ \tau_0 \le |t_1-t_2| \atop 0 \le t_1, t_2 <
  \infty} dt_1 dt_2 
  \sum_{a, c = 0}^{1} Q^{a c}_{r}(t_1,t_2) 
  \phi^{a}_{r}(t_1) \phi^{c}_{r}(t_2) \right.
  \nonumber \\
  & & \qquad - i \sum_{r} \int_{0}^{\infty} \! dt \, \left( 
    - 2 \, w_1 \, \phi^{0}_{r}(t) \phi^{1}_{r}(t) 
    - \sum_{p=2}^{\infty} 2 \, p \, w_p \, \phi^{0}_{r}(t)
    (\phi^{1}_r(t))^{2p-1} 
    \right)
  \nonumber \\
  & & \qquad - \frac{1}{2} \sum_{r, r'}  K_{r r'} 
  \int_{ |t_1-t_2| < \tau_0 \atop 0 \le t_1, t_2 < \infty} 
  dt_1 dt_2
  \sum_{a, c = 0}^{1}
  \phi^{a}_{r}(t_1) \phi^{c}_{r}(t_2) 
  \phi^{\bar{a}}_{r'}(t_1) \phi^{\bar{c}}_{r'}(t_2)
  \nonumber \\
  & & \qquad \left. 
  + i \sum_r \hat{\varphi_r} \left\{ \phi^{1}_r(0) - \varphi_r
  \right\}
  + \sum_r \int_{0}^{\infty} \! dt \, \left(
  \ell_r(t) \phi^{1}_r(t) + i h_r(t) \phi^{0}_r(t) 
  \right)
  \right\}.
  \label{eqa:S_fp_0}
\end{eqnarray}

In this form the fixed point action no longer contains the auxiliary
fields $Q^{a c}_r(t_1,t_2)$ for times $t_1,t_2$ such that $|t_1-t_2| <
\tau_0$. We now re-introduce those auxiliary fields through the same
Hubbard-Stratonovich transformation that was used to obtain
Eq.~(\ref{eq:Z0-mu}), and obtain:
\begin{eqnarray}
  S_{\mbox{\scriptsize fp}}[Q] & = & F_{\Omega} 
  + \frac{1}{2} \sum_{r, r'}  M_{r r'} 
  \int\!\!\int_{0}^{\infty} dt_1 dt_2 \sum_{a, c = 0}^{1} 
  Q^{a c}_{r}(t_1,t_2) Q^{\bar{a} \bar{c}}_{r'}(t_1,t_2) 
  \nonumber \\
  & & 
  - \ln \int {\cal D}\phi^{0} \, {\cal D}\phi^{1} \, 
  {\cal D}\hat{\varphi} \;
  \exp\left\{
  i \sum_{r} \int\!\!\int_{0}^{\infty} dt_1 dt_2 
  \sum_{a, c = 0}^{1} Q^{a c}_{r}(t_1,t_2) 
  \phi^{a}_{r}(t_1) \phi^{c}_{r}(t_2) \right.
  \nonumber \\
  & & \qquad - i \sum_{r} \int_{0}^{\infty} \! dt \, \left( 
    - 2 \, w_1 \, \phi^{0}_{r}(t) \phi^{1}_{r}(t) 
    - \sum_{p=2}^{\infty} 2 \, p \, w_p \, \phi^{0}_{r}(t)
    (\phi^{1}_r(t))^{2p-1} 
    \right)
  \nonumber \\
  & & \qquad \left. 
  + i \sum_r \hat{\varphi_r} \left\{ \phi^{1}_r(0) - \varphi_r \right\}
  + \sum_r \int_{0}^{\infty} \! dt \, \left(
  \ell_r(t) \phi^{1}_r(t) + i h_r(t) \phi^{0}_r(t) 
  \right)
  \right\}.
  \label{eqa:S_fp}
\end{eqnarray}

\section{Reparametrization Symmetry}
\label{sec:rpg}
We are now finally ready to evaluate the effect of a reparametrization
$t \to s(t)$ of the time variable on the fixed-point generating
functional $ {\cal Z}_{\mbox{\scriptsize fp}}[\ell, h]$. We consider
any smooth monotonous increasing function $s(t)$ satisfying the
boundary conditions $s(0)=0$ and $s(\infty)=\infty$, which induces the
following transformation of the sources:
\begin{eqnarray}
  \tilde{\ell}_{r}(t) & = &
  \left(\frac{\partial s}{\partial t}\right) \ell_{r}(s(t)) 
  \label{eqa:rpg-transf-sources-l} \\ 
  \tilde{h}_{r}(t) & = & h_{r}(s(t)), 
  \label{eqa:rpg-transf-sources-h}
\end{eqnarray}
and compute the fixed point disorder averaged generating functional,
evaluated at the transformed sources: 
\begin{eqnarray}
  {\cal Z}_{\mbox{\scriptsize fp}} 
  [\{\tilde{\ell}_r(t), \tilde{h}_r(t)\}] & = & 
  \int {\cal D}\tilde{Q} \; \exp\left( 
  - F_{\Omega} 
  - \frac{1}{2} \sum_{r, r'}  M_{r r'} 
  \int\!\!\int_{0}^{\infty} dt_1 dt_2 \sum_{a, c = 0}^{1} 
  \tilde{Q}^{a c}_{r}(t_1,t_2) \tilde{Q}^{\bar{a}
  \bar{c}}_{r'}(t_1,t_2) 
  \right. 
  \nonumber \\
  & & 
  + \ln \int {\cal D}\psi^{0} \, {\cal D}\psi^{1} \, 
  {\cal D}\hat{\varphi} \;
  \exp\left\{
  i \sum_{r} \int\!\!\int_{0}^{\infty} dt_1 dt_2 
  \sum_{a, c = 0}^{1} \tilde{Q}^{a c}_{r}(t_1,t_2) 
  \psi^{a}_{r}(t_1) \psi^{c}_{r}(t_2) \right.
  \nonumber \\
  & & \qquad - i \sum_{r} \int_{0}^{\infty} \! dt \, \left( 
    - 2 \, w_1 \, \psi^{0}_{r}(t) \psi^{1}_{r}(t) 
    - \sum_{p=2}^{\infty} 2 \, p \, w_p \, \psi^{0}_{r}(t)
    (\psi^{1}_r(t))^{2p-1} 
    \right)
  \nonumber \\
  & & \qquad 
  + i \sum_r \hat{\varphi_r} \left\{ \psi^{1}_r(0) - \varphi_r
  \right\} 
  \nonumber \\
  & & \qquad
  \left. \left. + \sum_r \int_{0}^{\infty} \! dt \, \left(
    \tilde{\ell}_r(t) \psi^{1}_r(t) + i \tilde{h}_r(t) \psi^{0}_r(t) 
    \right) \right\} \right).
  \label{eqa:S_fp_rpg0}
\end{eqnarray}
Here we have changed the name of the dummy variables from $\phi$ to
$\psi$ and from $Q$ to $\tilde{Q}$
in the functional integral. We now perform the changes of variables
$\psi^{a}_{r}(t) = \left(\frac{\partial s}{\partial
t}\right)^{\bar{a}} \phi^{a}_{r}(s(t)) $ and $\tilde{Q}^{a
c}_{r}(t_1,t_2) = \left(\frac{\partial s}{\partial t_1}\right)^{a}
\left(\frac{\partial s}{\partial t_2}\right)^{c} Q^{a
c}_{r}(s(t_1),s(t_2))$, i.e. the change of variables associated with
the RpG transformation of Eq.~(\ref{eq:rpg-transf}), thus obtaining:
\begin{eqnarray}
  \lefteqn{ {\cal Z}_{\mbox{\scriptsize fp}} 
  [\{\tilde{\ell}_r(t), \tilde{h}_r(t)\}] } 
  \nonumber \\
  & = & \int {\cal D}Q \; \exp\left(
  - F_{\Omega} 
  - \frac{1}{2} \sum_{r, r'}  M_{r r'} 
  \int\!\!\int_{0}^{\infty} dt_1 dt_2 \sum_{a, c = 0}^{1} 
  \left(\frac{\partial s}{\partial t_1}\right)^{a+\bar{a}} \!
  \left(\frac{\partial s}{\partial t_2}\right)^{c+\bar{c}} \!
  Q^{a c}_{r}(s(t_1),s(t_2)) \,
  Q^{\bar{a} \bar{c}}_{r'}(s(t_1),s(t_2))
  \right.
  \nonumber \\
  & & \qquad \qquad
  + \ln \int {\cal D}\phi^{0} \, {\cal D}\phi^{1} \, 
  {\cal D}\hat{\varphi} \;
  \exp\left\{ 
  i \sum_{r} \int\!\!\int_{0}^{\infty} dt_1 dt_2 
  \sum_{a, c = 0}^{1} 
  \left(\frac{\partial s}{\partial t_1}\right)^{a+\bar{a}} \! 
  \left(\frac{\partial s}{\partial t_2}\right)^{c+\bar{c}} \! 
  \right. \nonumber \\
  & & \qquad \qquad \qquad \qquad \qquad \qquad \qquad \qquad 
  \qquad \qquad \qquad \qquad \qquad \qquad 
  \times \,
  Q^{a c}_{r}(s(t_1),s(t_2)) \;
  \phi^{a}_{r}(s(t_1)) \;
  \phi^{c}_{r}(s(t_2))
  \nonumber \\[1.3ex]
  & & \qquad \qquad  \qquad - i \sum_{r} \int_{0}^{\infty} \! dt \, \left(
    - 2 \, w_1 \, 
    \left(\frac{\partial s}{\partial t}\right)
    \phi^{0}_{r}(s(t)) \, 
    \phi^{1}_{r}(s(t)) 
    - \sum_{p=2}^{\infty} 2 \, p \, w_p 
    \left(\frac{\partial s}{\partial t}\right)
    \phi^{0}_{r}(s(t)) \,
    (\phi^{1}_r(s(t)))^{2p-1} 
    \right)
  \nonumber \\
  & & \qquad \qquad  \qquad 
  + i \sum_r \hat{\varphi_r} \left\{ \phi^{1}_r(s(0)) - \varphi_r
  \right\} \nonumber \\
  & & \qquad \qquad  \qquad \left. + \sum_r \int_{0}^{\infty} \! dt \, \left\{ 
  \left(\frac{\partial s}{\partial t}\right) \ell_r(s(t)) \phi^{1}_r(s(t)) 
  + i h_r(s(t)) \left(\frac{\partial s}{\partial t}\right) \phi^{0}_r(s(t)) 
    \right\}
  + \ln {\cal J}_1 \left[\frac{{\cal D} \left( \psi^{0} \, \psi^{1} \right)} 
    {{\cal D} \left( \phi^{0} \, \phi^{1} \right)} \right] 
  \right\} 
  \nonumber \\
  & & \qquad \qquad
  \left. + \ln {\cal J}_2 \left[\frac{{\cal D} \tilde{Q}} 
    {{\cal D} Q } \right] \right)
  \label{eqa:S_fp_rpg1}
\end{eqnarray}
Here the symbol ${\cal J}_1 \left[\frac{{\cal D} \left( \psi^{0} \,
\psi^{1} \right) } {{\cal D} \left( \phi^{0} \, \phi^{1} \right) }
\right] $ represents the Jacobian of the transformation from $\phi$ to
$\psi$, and the symbol ${\cal J}_2 \left[\frac{{\cal D} \tilde{Q}}
{{\cal D} Q } \right]$ represents the Jacobian of the transformation
from $Q$ to $\tilde{Q}$.  Since both transformations are linear
transformations, the Jacobians only depend on the reparametrization
$s(t)$, but they {\em do not} depend on the fields $\phi$ or $Q$, or
the sources $\{ \ell_r(t), h_r(t)\}$. For this reason, we will denote
them as ${\cal J}_1\{s(t)\}$ and ${\cal J}_2\{s(t)\}$ respectively.
Using the fact that $a+\bar{a}=1$ and $c+\bar{c}=1$, we find that the
factor $\left(\frac{\partial s}{\partial t_1}\right)^{a+\bar{a}}
\left(\frac{\partial s}{\partial t_2}\right)^{c+\bar{c}}$ is simply
the Jacobian of the transformation from $(t_1,t_2)$ to
$(t'_1,t'_2)=(s(t_1),s(t_2))$, and therefore we obtain:
\begin{eqnarray}
  \lefteqn{ {\cal Z}_{\mbox{\scriptsize fp}} 
  [\{\tilde{\ell}_r(t), \tilde{h}_r(t)\}] }
  \nonumber \\
  & = & \int {\cal D}Q \; \exp\left(
  - F_{\Omega} 
  - \frac{1}{2} \sum_{r, r'}  M_{r r'} 
  \int\!\!\int_{0}^{\infty} dt'_1 dt'_2 \sum_{a, c = 0}^{1} 
  Q^{a c}_{r}(t'_1,t'_2) \,
  Q^{\bar{a} \bar{c}}_{r'}(t'_1,t'_2)
  \right.
  \nonumber \\
  & & \qquad \qquad
  + \ln \int {\cal D}\phi^{0} \, {\cal D}\phi^{1} \, 
  {\cal D}\hat{\varphi} \;
  \exp\left\{ 
  i \sum_{r} \int\!\!\int_{0}^{\infty} dt'_1 dt'_2 
  \sum_{a, c = 0}^{1} 
  Q^{a c}_{r}(t'_1,t'_2) \;
  \phi^{a}_{r}(t'_1) \;
  \phi^{c}_{r}(t'_2)
  \right. 
  \nonumber \\
  & & \qquad \qquad  \qquad - i \sum_{r} \int_{0}^{\infty} \! dt' \, \left( 
    - 2 \, w_1 \, 
    \phi^{0}_{r}(t') \, 
    \phi^{1}_{r}(t') 
    - \sum_{p=2}^{\infty} 2 \, p \, w_p 
    \phi^{0}_{r}(t') \,
    (\phi^{1}_r(t'))^{2p-1} 
    \right)
  \nonumber \\
  & & \qquad \qquad  \qquad 
  + i \sum_r \hat{\varphi_r} \left\{ \phi^{1}_r(0) - \varphi_r
  \right\} \nonumber \\
  & & \qquad \qquad  \qquad \left. + 
  \sum_r \int_{0}^{\infty} \! dt' \, \left(
  \ell_r(t') \phi^{1}_r(t') 
  + i h_r(t') \phi^{0}_r(t') 
    \right)
  + \ln {\cal J}_1\{s(t)\}
  \right\} 
  \nonumber \\
  & & \qquad \qquad
  \left. + \ln {\cal J}_2\{s(t)\} \right)
  \nonumber \\[2.0ex]
  & = & {\cal Z}_{\mbox{\scriptsize fp}} 
  [\{ \ell_r(t), h_r(t)\}]
  \times {\cal J}_1\{s(t)\}\times {\cal J}_2\{s(t)\} 
  \label{eqa:S_fp_rpg}
\end{eqnarray}
Here we have used the boundary condition $s(0)=0$. We now consider the
special case of zero sources, i.e. $\ell_r(t)=0$, $h_r(t)=0$; in this
case, the transformed sources are identical to the original ones, and
we have the condition
\begin{equation}
  {\cal Z}_{\mbox{\scriptsize fp}} 
  [\{ 0, 0\}] 
  = {\cal Z}_{\mbox{\scriptsize fp}} 
  [\{ 0, 0\}] 
  \times {\cal J}_1\{s(t)\}\times {\cal J}_2\{s(t)\}. 
  \label{eq:Z-zero-sources}
\end{equation}
Since the generating functional is nonzero for zero sources (it is
actually unity~\cite{MSR-functional-integral_dedominicis-peliti}), we
immediately conclude that, for {\em any} reparametrization $s(t)$ the
product of the Jacobians is unity: ${\cal J}_1\{s(t)\}\times {\cal
J}_2\{s(t)\} \equiv 1$. Thus we obtain, for {\em any}
reparametrization $s(t)$, the identity:
\begin{equation}
  {\cal Z}_{\mbox{\scriptsize fp}} 
  [\{\tilde{\ell}_r(t), \tilde{h}_r(t)\}] 
  = {\cal Z}_{\mbox{\scriptsize fp}} 
  [\{ \ell_r(t), h_r(t)\}], 
  \label{eq:Z-invariant-reparametrization}
\end{equation}
i.e., we have shown that the fixed-point generating functional is {\em
invariant} under time reparametrization transformations.

\section{Physical consequences of the time reparametrization symmetry}
\label{sec:consequences}

Since the renormalization group procedure described above involves 
integrating over all short time-scale fluctuations, the fixed point
generating functional that we obtained controls the long-time
dynamics of the model. The group of transformations associated with
time reparametrizations is a continuous symmetry group for the fixed
point generating functional. This symmetry is broken by the actual
dynamical correlations and responses observed in the system. As an
example, let us consider the space-averaged two-time correlation
$C(t,t_w) \equiv \frac{1}{N} \sum_r \langle \phi_{r}(t)
\phi_{r}(t_w)\rangle $, where $t_w$ is normally referred to as the
``waiting time'' and $t$ as the ``final time''. If this correlation
was actually invariant under time reparametrizations, we would have
$C(t,t_w) = C(s(t), s(t_w))$ for any arbitrary increasing function
$s(t)$ such that $s(0)=0$ and $s(\infty) = \infty$. The only possible
way that this condition can be satisfied is if $C(t,t_w) = C_0$ (a
constant). Since correlations in spin glasses actually {\em do} change
with time, this implies that the reparametrization symmetry must be
broken.

We have therefore the presence of a broken continuous symmetry
group. Since no long range interactions or gauge potentials are
present, we should normally expect that a Goldstone theorem applies,
giving rise to the presence of Goldstone modes (or soft modes) in the
system~\cite{Amit}. For this reason, it has already been argued in
Refs.~\cite{ckcc_short-rpg_prl-2002, ccck-prl-numeric, cccki-prb,
chamon-charb-cug-reich-sellito_condmat04} that Goldstone modes should
be present in the non-equilibrium dynamics of spin glasses and
possibly other glassy systems, and could in principle constitute the
main source of fluctuations in the non-equilibrium dynamics of these
systems. In other words, the presence of time reparametrization
symmetry could account for a significant part of the dynamical
heterogeneity effects observed in glassy systems. 

In general, Goldstone modes are obtained from a continuous symmetry
transformation by making it smoothly space dependent. For example, if
the symmetry corresponds to spin rotations by any angle, to obtain a
Goldstone mode the angle is chosen to be smoothly space dependent. In
the present case, the continuous symmetry corresponds to
reparametrizing the time variable $t \to s(t)$. In the uniform case,
this leads to the symmetry transformation $C(t,t_w) \to
\tilde{C}(t,t_w) = C(s(t),s(t_w))$. The Goldstone modes are obtained
by choosing the time reparametrization to be smoothly space dependent,
i.e. $t \to s_r(t)$, and $C_r(t,t_w) = C_0(s_r(t),s_r(t_w))$, where
$C_0(t,t_w)$ is space independent~\cite{ ccck-prl-numeric, cccki-prb,
chamon-charb-cug-reich-sellito_condmat04}. Since the reparametrization
is non-uniform, it is no longer a symmetry transformation for the
system, but if the space variation is slow enough, the change in the
action with respect to the value for a uniform two-time field is
small. A possible (very simplified) physical interpretation
of these Goldstone modes, is that they are associated with
``non-uniform slow relaxation'': if one considers different small
regions in the system, for all regions the relaxation path is very
nearly the same (as given by $C_0(t,t_w)$), but the rate at which each
small region advances in its relaxation path can fluctuate from region
to region.

Testing for the presence of fluctuations associated with this
reparametrization symmetry has been performed in numerical simulations
of both spin glasses and structural glasses. In spin glasses, one
prediction that can be tested in simulations refers to the values of
coarse grained {\em local} correlations $ C_r(t,t_w) \equiv
\frac{1}{n} \sum_{i\in B_r} s_i(t) s_i(t_w) $ and integrated responses
$\chi_r(t,t_w) \equiv \int_{t_w}^{t} dt' \frac{1}{n} \sum_{i\in B_r}
\frac { \delta \langle \phi_{i}(t) \rangle }{ \delta h_{i}(t') } $;
where $B_r$ is a cubic coarse graining box containing $n$ spins
centered at the point $r$; in comparison to the {\em global} values
$C(t,t_w)$ and $\chi (t,t_w)$ obtained by taking the averages over the
whole sample. As explained in Refs.~\cite{ccck-prl-numeric,cccki-prb},
the presence of Goldstone modes associated with time reparametrization
symmetry would imply that the pairs $(C_r,\chi_r)$ should be
concentrated predominantly along the parametric curve $\chi(C)$. It
turns out that this is exactly what is observed in the results of
numerical simulations in the 3D Edwards-Anderson
model~\cite{ccck-prl-numeric,cccki-prb}. Another testable prediction
is that, if the global correlation $C(t,t_w)$ is only a function of
the ratio $t/t_w$, i.e. $C(t,t_w)={\cal C}(t/t_w)$ , the probability
distribution $\rho(C_r(t,t_w))$ for the values of the local coarse
grained correlation $C_r(t,t_w)$ should collapse as a function of
$t_w$, as long as $t/t_w$ is held fixed. This has also been found to
be the case in simulations in the 3D Edwards-Anderson
model~\cite{ccck-prl-numeric,cccki-prb}. In
Ref.~\cite{chamon-charb-cug-reich-sellito_condmat04} a more detailed
study of the shape of the probability distributions for both the
Edwards-Anderson model and a kinetically constrained model of
glassiness was performed, with results that were consistent with the
predictions derived from the presence of Goldstone modes in the
system.

Another aspect of the results presented here that can be tested by
comparison with numerical simulations in spin glasses is the fact that
the symmetry is only exact for the fixed point generating functional,
i.e. in the limit $t \to \infty$. For the case of an exact continuous
symmetry, one should expect that the presence of a true Goldstone mode
(with zero mass) gives rise to spatial correlations that decay as
power laws at long distances. 
However, at any finite time the symmetry is broken by small
corrections to the action, which we can think of as small
symmetry-breaking fields that go to zero at $t \to \infty$. As a
consequence of the presence of these symmetry breaking fields, the
Goldstone modes now acquire a small mass, which should vanish in the
$t \to \infty$ limit. In~Ref.~\cite{cccki-prb}, the spatial
correlation length $\xi(t,t_w)$ for fluctuations of the quantity
$Q^{11}_r(t,t_w)$ was measured in large scale long time simulations in
a 3D Edwards-Anderson model. For very large $t$, $t_w$ and $t/t_w$ the
time dependence of $\xi(t,t_w)$ was found to be consistent both with a
form $\xi(t,t_w) \approx \ln(t t_w)$ or a form $\xi(t,t_w) \approx (t
t_w)^a$, with $a \approx 0.04$. Both forms extrapolate (albeit slowly)
to infinity at infinite times. This is suggestive, and consistent with
what is expected from the results of the present work, but the actual
values of $\xi(t,t_w)$ are too small to make any firm statements about
the $t \to \infty$ limit.

The present work only proves the presence of time reparametrization
invariance in spin glasses. However, it is conceivable that the
symmetry could extend to structural glasses, and there has already
been some work in structural glasses which has found suggestive
evidence for its presence. In the case of structural glasses, there is
another quantity which plays the role of local coarse grained
correlation. It is defined~\cite{parsaeian-castillo_aging-hetdyn} as
$C_{{\bf r}}(t, t_w) = \frac{1}{N(B_{\bf r})} \sum_{{\bf r}_j(t_w) \in
  B_{\bf r}} \cos({\bf q} \cdot ({\bf r}_j(t) - {\bf r}_j(t_w)))$.
Here $B_{\bf r}$ denotes a coarse graining box centered at the point
${\bf r}$ in the system, and the sums run over all of the $N(B_{\bf r})$
particles present in $B_{\bf r}$ at the waiting time $t_w$.  The value
of q is usually chosen to correspond to the main peak in the structure
factor $S(q)$ of the system.  Unlike in the 3D Edwards-Anderson model,
in structural glasses the global correlation $C(t,t_w)$ is not a
function of the ratio $t/t_w$. For this situation, the presence of the
Goldstone mode associated with time reparametrization invariance
implicates that the probability distribution $\rho(C_r(t,t_w))$ for
the values of the local coarse grained correlation $C_r(t,t_w)$ should
collapse as a function of $t_w$, as long as the global correlation
$C(t,t_w)$ is held fixed~\cite{cccki-prb,
  chamon-charb-cug-reich-sellito_condmat04}. This has been found to be
the case, to a good approximation, in simulations in binary
Lennard-Jones mixtures and binary Weeks-Chandler-Anderson
mixtures~\cite{parsaeian-castillo_aging-hetdyn}.

Confocal microscopy experiments in colloidal
glasses~\cite{Weeks-Weitz, Weeks-Weitz_prl02,
  Courtland-Weeks-jphysc03} provide detailed data that include the
positions of all colloidal particles in some sub-volume of the sample
at different times in the evolution of the system. These data can be
analyzed in completely analogous ways to those used to analyze data
from simulations in structural glasses. It remains an open question
whether or not such analysis would provide further evidence in favor
of the presence of time reparametrization symmetry.

\section{Summary}
\label{sec:summary}

In this work, we have presented a detailed proof of the
presence of a symmetry under continuous reparametrizations of the time
variable, for the long time dynamics of a generic spin glass model
with two-spin interactions. No assumptions were made about the range of
the interactions, therefore the proof applies equally to short-range
models, such as the Edwards-Anderson model, and to long-range models,
such as the Sherrington-Kirkpatrick model. By performing a
Renormalization Group procedure that exactly integrates over degrees
of freedom associated with {\em short time differences}, we have
obtained the RG flow for the parameters in the action. We have found
that the RG flow converges to a fixed-point generating functional, and
we have explicitly written the form of this generating functional. Our
main result is to have shown that the value of the fixed point
generating functional is left invariant by a transformation of the
sources induced by a monotonous increasing but otherwise arbitrary
reparametrization of the time variable.

The group of transformations associated with time reparametrizations
is a continuous symmetry group for the fixed point generating
functional. This symmetry is broken by the actual dynamical
correlations and responses observed in the system. In a situation like
this, one would normally expect the presence of Goldstone
modes. Indeed, it has been argued~\cite{ckcc_short-rpg_prl-2002,
  ccck-prl-numeric, cccki-prb,
  chamon-charb-cug-reich-sellito_condmat04} that Goldstone modes
associated with time reparametrization invariance should dominate the
fluctuations in the non-equilibrium dynamics of these
systems. Positive evidence for this statement has been found in
simulations of the aging dynamics of the 3D Edwards-Anderson
model~\cite{ccck-prl-numeric, cccki-prb,
  chamon-charb-cug-reich-sellito_condmat04}. Even simulations in
systems without quenched disorder, such as kinetically constrained
models of glassiness~\cite{chamon-charb-cug-reich-sellito_condmat04}
and models of structural
glasses~\cite{parsaeian-castillo_aging-hetdyn} show evidence in favor
of the presence of this symmetry. Additionally, experimental tests for
the presence of this symmetry in colloidal glasses can be provided by
confocal microscopy measurements. 
Having proved the presence of time reparametrization symmetry, the
present work opens the door for a more detailed analytical study of
the symmetry itself, of the Goldstone modes probably associated with
its presence, and more generally of the fluctuations (``dynamical
heterogeneities'') that are present
in the slow dynamics of spin glasses and other glassy systems.

\section{Acknowledgments}
\label{sec:ack}

The author wishes to especially thank C.~Chamon, L.~Cugliandolo
and M.~Kennett for very enlightening discussions over the years, and
J.~P.~Bouchaud, D.~Reichman, and G.~Biroli for suggestions and
discussions.  This work was supported in part by DOE under grant
DE-FG02-06ER46300, by NSF under grant PHY99-07949, and by Ohio
University. Part of this work was completed during a stay at the Aspen
Center for Physics, and the author would like to thank the Center for
its hospitality.

\end{document}